\documentclass[prd,tightenlines,floatfix,showpacs,
nofootinbib,eqsecnum]{revtex4}
\usepackage[dvips,final]{graphicx}
\usepackage{epsfig}
\begin{document}

\date{\today}
\title{Higher twist distribution amplitudes of the pion and electromagnetic
form factor $F_{\pi}(Q^2)$ }
\author{S. S. Agaev }
\email{agaev_shahin@yahoo.com}
\affiliation{Institute for Physical Problems, Baku State University,\\
Z. Khalilov st.\ 23, Az-1148 Baku,
Azerbaijan}

\begin{abstract}
The pion electromagnetic form factor is calculated within the QCD light-cone
sum rule method and using a renormalon model for the higher twist
distribution amplitudes (DAs). The theoretical predictions are compared with
the experimental data and constraints on the pion leading and twist-4 DAs
are extracted. An upper bound on the twist-4 contribution to the form factor
and estimates of effects due to higher conformal spins in the pion twist-4
DAs are obtained.
\end{abstract}

\maketitle

\section{Introduction}

The leading and higher twist distribution amplitudes (DAs) of hadrons are
important ingredients in investigation of various exclusive processes within
QCD \cite{ER80,LB80}. The leading twist DAs appear in the QCD factorization
formulas and describe exclusive processes with the leading power accuracy.
They correspond to parton configurations of hadrons with a minimal number of
constituents. The higher twist DAs are essential in computing different
power-suppressed corrections, which emerge due to parton virtuality,
transverse momentum, contributions of higher Fock states with a nonminimal
number of hadron constituents.

The traditional method for the description of DA is founded on the conformal
symmetry of the QCD Lagrangian \cite{BKM03}. Within this approach the
leading and higher twist DAs are expanded over the conformal spin. It is
important that any parametrization of DA based on a truncated conformal
expansion is consistent with the QCD equations of motion (EOM) \cite{BF90}
and is preserved by the QCD evolution to the leading logarithmic accuracy %
\cite{ER80,LB80}. Therefore, the conformal expansion provides a practical
framework for modeling of hadron DAs \cite{BF90,CZ84,BB98,BB99,Ball99,BB03}
and is widely used for investigation of numerous exclusive processes in QCD.

Because of the increasing number of parameters at higher conformal spins
and practical difficulties in phenomenological applications, one has to
restrict one's self by only the first few terms in the conformal expansion of
DAs. As a result, the contributions of higher conformal spins to DAs in the
existing calculations are
neglected. At the same time, the suppression of higher spin contributions
and the convergence of conformal expansion at present experimentally
accessible energy regimes is by no means obvious and may be wrong.
Therefore, one needs to draw new approaches to clarify this problem.

The renormalon model proposed recently in Refs.\ \cite{An00,BGG04} pursues to
test precisely this issue, that is to set a plausible upper bound for the
possible contributions of higher conformal spins that so far escaped
attention. The renormalon approach employs the assumption that the infrared
(IR) renormalons in the leading twist coefficient functions should cancel
the ultraviolet (UV) renormalons in the matrix elements of twist-4 operators
in a relevant operator product expansion. Such cancellation was proved by
explicit calculations in the case of the simple exclusive amplitude
involving pseudoscalar and vector mesons \cite{BGG04}. It turned out that
this is enough to obtain the full set of two- and three-particle twist-4 DAs
of pseudoscalar and vector mesons in terms of the leading twist DAs. It is
remarkable that the set of twist-4 DAs depend only on one new parameter,
which can be related to the matrix element of some local operator (see Sec.\ II )
 and estimated using the QCD sum rule. In other words, the twist-2 and
twist-4 DAs of pseudoscalar and vector mesons can be determined using the
same set of parameters that considerably restricts a freedom in the choice
of DAs, increasing, at the same time, the predictive power and reliability
of QCD\ results.

The renormalon calculus was employed in Ref.\ \cite{BGG04}, where the pion
and $\rho $-meson twist-4 DAs were constructed. In the calculations the
mesons asymptotic DAs were used. A generic feature of the renormalon model is
that it predicts higher twist distributions that are larger at the
end points compared to the asymptotic distributions, and are expected to
give rise to larger higher twist effects in exclusive reactions. The main
purpose of this work is to test this idea on example of the pion
electromagnetic form factor (FF), that is to set an upper bound on possible
twist-4 contributions to FF. To this end, we extend results of Ref.\ \cite%
{BGG04} and compute the pion higher twist DAs using the leading twist DA
with two nonasymptotic terms. We apply our predictions for studying the pion
form factor within the QCD light-cone sum rule (LCSR) method and extract
constraints on the input parameters $b_{2}(\mu _{0}^{2})$ and $b_{4}(\mu
_{0}^{2})$ at the normalization scale $\mu _{0}^{2}=1\,\,\rm{GeV}^{2}$.

This paper is structured as follows: In Sec.\ II we define the two-- and
three-particle twist-4 DAs of the pion and calculate them within the
renormalon approach. In Sec.\ III general expressions for the FF $F_{\pi
}(Q^{2})$ in the context of the QCD LCSR method with twist-6 accuracy are
presented. In Sec.\ IV we confront our predictions with the available data on 
$F_{\pi }(Q^{2})$ and by this way model the pion DAs. Section\ V is reserved for
concluding remarks.

\section{Higher twist DAs of the pion}

The light-cone two-particle distribution amplitudes of the pion are defined
through the light-cone expansion of the matrix element,
\[
\left\langle 0\left| \overline{d}(x_{2})\gamma _{\nu }\gamma _{5}\left[
x_{2},x_{1}\right] u(x_{1})\right| \pi ^{+}(p)\right\rangle = 
\]
\[
=if_{\pi }p_{\nu }\int_{0}^{1}due^{-iupx_{1}-i\overline{u}px_{2}}\left[
\varphi ^{(2)}(u,\,\mu _{F}^{2})+\Delta ^{2}\varphi _{1}^{(4)}(u,\,\,\mu
_{F}^{2})+O(\Delta ^{4})\right] 
\]
\begin{equation}
+if_{\pi }\left( \Delta _{\nu }(p\Delta )-p_{\nu }\Delta ^{2}\right)
\int_{0}^{1}due^{-iupx_{1}-i\overline{u}px_{2}}\left[ \varphi
_{2}^{(4)}(u,\,\,\mu _{F}^{2})+O(\Delta ^{4})\right] ,  \label{eq:1}
\end{equation}%
where $\varphi ^{(2)}(u,\,\mu _{F}^{2})=\varphi _{\pi }(u,\,\mu _{F}^{2})$
is the leading twist DA of the pion, whereas $\varphi _{1}^{(4)}(u,\,\,\mu
_{F}^{2}),\,\,\varphi _{2}^{(4)}(u,\,\,\mu _{F}^{2})$ are two-particle
twist-4 DAs. We use the notation $\left[ x_{2},x_{1}\right] $ for the Wilson
line connecting the points $x_{1}$ and $x_{2}$,
\begin{equation}
\left[ x_{2},x_{1}\right] =P\exp \left[ -ig\int_{0}^{1}dt\Delta _{\mu
}A^{\mu }(x_{2}+t\Delta )\right] .  \label{eq:2}
\end{equation}
In Eqs.\ (\ref{eq:1}) and (\ref{eq:2}) $\Delta =x_{1}-x_{2}$ and$\,\,
\overline{u}=1-u$.

Apart from the two-particle DAs there exist the three-particle twist-4 DAs
involving an extra gluon field, which we define in the form \cite{BF90,BGG04}
\[
\left\langle 0\left| \overline{d}(-z)\left[ -z,vz\right] \gamma _{\nu
}\gamma _{5}gG_{\mu \rho }(vz)\left[ vz,z\right] u(z)\right| \pi
^{+}(p)\right\rangle 
\]
\[
=f_{\pi }\int D\alpha _{i}e^{-ipz(\alpha _{1}-\alpha _{2}+\alpha
_{3}v)}\left\{ \frac{p_{\nu }}{pz}(p_{\mu }z_{\rho }-p_{\rho }z_{\mu })\Phi
_{\parallel }(\alpha _{1},\alpha _{2},\alpha _{3})\right. 
\]
\begin{equation}
\left. +\left[ p_{\rho }\left( g_{\mu \nu }-\frac{z_{\mu }p_{\nu }}{pz}
\right) -p_{\mu }\left( g_{\rho \nu }-\frac{z_{\rho }p_{\nu }}{pz}\right) 
\right] \Phi _{\perp }(\alpha _{1},\alpha _{2},\alpha _{3})\right\} ,
\label{eq:3}
\end{equation}
where the longitudinal momentum fraction of the gluon is $\alpha _{3}$ and
the integration measure is defined as 
\begin{equation}
\int D\alpha _{i}=\int_{0}^{1}d\alpha _{1}d\alpha _{2}d\alpha _{3}\delta
(1-\alpha _{1}-\alpha _{2}-\alpha _{3}).  \label{eq:4}
\end{equation}
The other pair of DAs that can be obtained from Eq.\ (\ref{eq:3}) after the
replacement $\gamma _{5}G_{\mu \rho }\to i\widetilde{G}^{\mu \rho }=
\frac{i}{2}\epsilon ^{\mu \rho \alpha \beta }G_{\alpha \beta },$ 
\[
\left\langle 0\left| \overline{d}(-z)\left[ -z,vz\right] \gamma _{\nu }ig
\widetilde{G}_{\mu \rho }(vz)\left[ vz,z\right] u(z)\right| \pi
^{+}(p)\right\rangle 
\]%
\[
=f_{\pi }\int D\alpha _{i}e^{-ipz(\alpha _{1}-\alpha _{2}+\alpha
_{3}v)}\left\{ \frac{p_{\nu }}{pz}(p_{\mu }z_{\rho }-p_{\rho }z_{\mu })\Psi
_{\parallel }(\alpha _{1},\alpha _{2},\alpha _{3})\right. 
\]%
\begin{equation}
\left. +\left[ p_{\rho }\left( g_{\mu \nu }-\frac{z_{\mu }p_{\nu }}{pz}
\right) -p_{\mu }\left( g_{\rho \nu }-\frac{z_{\rho }p_{\nu }}{pz}\right) 
\right] \Psi _{\perp }(\alpha _{1},\alpha _{2},\alpha _{3})\right\}.
\label{eq:5}
\end{equation}
One more three-particle DA $\Xi _{\pi }(\alpha _{i})$ is introduced through
the formula \cite{BGG04}
\[
\left\langle 0\left| \overline{d}(-z)\left[ -z,vz\right] \gamma _{\nu
}\gamma _{5}gD^{\alpha }G_{\alpha \rho }(vz)\left[ vz,z\right] u(z)\right|
\pi ^{+}(p)\right\rangle 
\]
\begin{equation}
=if_{\pi }p_{\nu }p_{\rho }\int D\alpha _{i}e^{-ipz(\alpha _{1}-\alpha
_{2}+\alpha _{3}v)}\Xi _{\pi }(\alpha _{1},\alpha _{2},\alpha _{3}).
\label{eq:6}
\end{equation}
In this work we do not consider twist-4 four quark operators and
corresponding DAs.

Because of the EOM the two-particle DAs $\varphi _{1}^{(4)}(u,\,\,\mu
_{F}^{2}),\,\,\varphi _{2}^{(4)}(u,\,\,\mu _{F}^{2})$ can be expressed in
terms of the three-particle ones. Namely, from exact operator identities 
\cite{Br89}, which relate integrals over $v$ of the quark-gluon-antiquark
operator in Eq. (\ref{eq:3}) to derivatives of the quark-antiquark operator 
(\ref{eq:1}), it follows that
\[
\varphi _{2}^{(4)}(u)=\int_{0}^{u}dv\int_{0}^{v}d\alpha
_{1}\int_{0}^{1-v}d\alpha _{2}\frac{1}{\alpha _{3}}\left[ 2\Phi _{\perp
}-\Phi _{\parallel }\right] (\alpha _{1},\alpha _{2},\alpha _{3}),
\]%
\begin{equation}
\varphi _{1}^{(4)}(u)+\varphi _{2}^{(4)}(u)=\frac{1}{2}\int_{0}^{u}d\alpha
_{1}\int_{0}^{1-u}d\alpha _{2}\frac{\overline{u}\alpha _{1}-u\alpha _{2}}{%
\alpha _{3}^{2}}\left[ 2\Phi _{\perp }-\Phi _{\parallel }\right] (\alpha
_{1},\alpha _{2},\alpha _{3}),  \label{eq:7}
\end{equation}%
where $\alpha _{3}=1-\alpha _{1}-\alpha _{2}$.

The standard method to handle meson DAs is modeling them employing the
conformal expansion. Then for the leading twist pion DA we get 
\begin{equation}
\varphi _{\pi }(u,\,\mu _{F}^{2})=\varphi _{asy}(u)\left[ 1+\sum_{n=2,4..}^{
\infty }b_{n}(\mu _{F}^{2})C_{n}^{3/2}(u-\overline{u}) \right].  \label{eq:8}
\end{equation}
Here $\varphi _{asy}(u)$ is the pion asymptotic DA
\[
\varphi _{asy}(u)=6u\overline{u}, 
\]
and $C_{n}^{3/2}(\xi )$ are the Gegenbauer polynomials. The functions 
$b_{n}(\mu _{F}^{2})$ determine the evolution of $\varphi _{\pi }(u,\,\mu
_{F}^{2})$ on the factorization scale $\,\mu _{F}^{2}$,
\begin{equation}
b_{n}(\mu _{F}^{2})=b_{n}(\mu _{0}^{2})\left[ \frac{\alpha _{\rm{S}}(\mu
_{F}^{2})}{\alpha _{\rm{S}}(\mu _{0}^{2})}\right] ^{\gamma _{n}/\beta
_{0}},\,\,\,\,\gamma _{n}=C_{F}\left[ 1-\frac{2}{(n+1)(n+2)}%
+4\sum_{j=2}^{n+1}\frac{1}{j}\right] .  \label{eq:9}
\end{equation}
In the above, $\gamma _{n}$ are the anomalous dimensions and $\mu
_{0}^{2} $ is the normalization scale. The expansion over the conformal spin 
$j$ can also be performed for the higher twist DAs \cite
{BB99,Ball99,BK99,BK01}.

The renormalon approach to the higher twist DAs is based on another idea.
To explain principle points of the renormalon approach and derive relations
between the pion twist two and four DAs in Ref.\ \cite{BGG04}, the authors
considered the gauge-invariant time-ordered product of two quark currents at
a small light-cone separation,

\[
\left\langle 0\left| T\{ \overline{d}(x_{2})\gamma _{\nu
}\gamma _{5}\left[ x_{2},x_{1}\right] u(x_{1})\}\right| \pi
^{+}(p)\right\rangle 
\]

\begin{equation}
=if_{\pi }\int_{0}^{1}due^{-iupx_{1}-i\overline{u}px_{2}}\left[
G_{1}(u,\Delta ^{2},\,\mu ^{2})p_{\nu }+G_{2}(u,\Delta ^{2},\,\mu
^{2})\left( \frac{p\Delta }{\Delta ^{2}}\Delta _{\nu }-p_{\nu }\right) 
\right] ,  \label{eq:9a}
\end{equation}
with $\left| \Delta ^{2}\right| \ll 1/\Lambda ^{2}$ and $\Delta ^{2}<0$
playing the role of the hard scale and $\mu ^{2}$ being the ultraviolet
renormalization scale. This martix element is parametrized in terms of two
Lorentz-invariant amplitudes $G_{1}(u,\Delta ^{2},\,\mu ^{2})$ and $%
G_{2}(u,\Delta ^{2},\,\mu ^{2})$, which in the light-cone limit $\Delta
^{2}\to 0$ with $p\Delta $ fixed have the expansions%
\begin{equation}
G_{i}(u,\Delta ^{2})=C_{i}^{(2)}\otimes \varphi ^{(2)}+\Delta ^{2}\sum
C_{i}^{(4)}\otimes \varphi _{i}^{(4)}+O(\Delta ^{4}).  \label{eq:9b}
\end{equation}
Here $C_{i}^{(t)}$ are the coefficient functions and $\varphi ^{(t)}$ are
the pion DAs, $t$ refers to twist, and the summation runs over all
contributions for a given twist.

Considering in (\ref{eq:9b}) the leading twist coefficient functions to all
orders of $\alpha _{\rm{S}}(\mu ^{2})$ and the twist-4 contribution to
the leading order one gets
\[
G_{1}(u,\Delta ^{2})=\left[ 1+c_{1}\alpha _{\rm{S}}+c_{2}\alpha _{
\rm{S}}^{2}+....\right] \otimes \varphi ^{(2)}+\Delta ^{2}\varphi
_{1}^{(4)}(u,\mu _{F}^{2}), 
\]%
\begin{equation}
G_{2}(u,\Delta ^{2})=\left[ \widetilde{c}_{1}\alpha _{\rm{S}}+\widetilde{%
c}_{2}\alpha _{\rm{S}}^{2}+....\right] \otimes \varphi ^{(2)}+\Delta
^{2}\varphi _{2}^{(4)}(u,\mu _{F}^{2}).  \label{eq:9c}
\end{equation}
Calculation of the leading twist coefficient functions to all orders using
the running coupling method gives rise to IR renormalon ambiguities in the
amplitudes $G_{i}(u,\Delta ^{2})$. These ambiguities are expressable in
terms of the pion leading twist DA. The twist-4 DAs (\ref{eq:3}),
(\ref{eq:5}), and (\ref{eq:6}) 
contain UV renormalon divergences, which were employed in Ref.\ \cite{BGG04}
to compute UV renormalon ambiguities in the pion two-particle twist-4 DAs 
$\varphi _{i}^{(4)}(u,\mu _{F}^{2})$. These UV renormalon ambiguities cancel
the IR renormalon ones in the sum of the different twists (\ref{eq:9c}), in
the same way as the logarithmic scale dependence is cancelled between matrix
elements and coefficient functions for a given twist. As a result, the
structure functions $G_{i}(u,\Delta ^{2})$ are unambiguous to the twist-4
accuracy. The idea of the renormalon model for the pion twist-4 DAs is to
define them by taking the functional form of the corresponding UV renormalon
ambiguities and replacing the overall normalization constant by a suitable
nonperturbative parameter. By this way one obtains the following relations
between the DAs of the pion:

\[
\Phi _{\perp }(\alpha _{1},\alpha _{2},\alpha _{3})=\frac{\delta ^{2}}{6}%
\left[ \frac{\varphi _{\pi }(\alpha _{1})}{1-\alpha _{1}}-\frac{\varphi
_{\pi }(\alpha _{2})}{1-\alpha _{2}}\right],
\]
\[
\Phi _{\parallel }(\alpha _{1},\alpha _{2},\alpha _{3})=\frac{\delta ^{2}}{3}
\left[ \frac{\alpha _{2}\varphi _{\pi }(\alpha _{1})}{(1-\alpha _{1})^{2}}-
\frac{\alpha _{1}\varphi _{\pi }(\alpha _{2})}{(1-\alpha _{2})^{2}}\right],
\]
\[
\Psi _{\perp }(\alpha _{1},\alpha _{2},\alpha _{3})=\frac{\delta ^{2}}{6}%
\left[ \frac{\varphi _{\pi }(\alpha _{1})}{1-\alpha _{1}}+\frac{\varphi
_{\pi }(\alpha _{2})}{1-\alpha _{2}}\right],
\]
\[
\Psi _{\parallel }(\alpha _{1},\alpha _{2},\alpha _{3})=-\frac{\delta ^{2}}{3
}\left[ \frac{\alpha _{2}\varphi _{\pi }(\alpha _{1})}{(1-\alpha _{1})^{2}}+
\frac{\alpha _{1}\varphi _{\pi }(\alpha _{2})}{(1-\alpha _{2})^{2}}\right],
\]
\begin{equation}
\Xi _{\pi }(\alpha _{1},\alpha _{2},\alpha _{3})=-\frac{2\delta ^{2}}{3}%
\left[ \frac{\alpha _{2}\varphi _{\pi }(\alpha _{1})}{1-\alpha _{1}}-\frac{%
\alpha _{1}\varphi _{\pi }(\alpha _{2})}{1-\alpha _{2}}\right].
\label{eq:10}
\end{equation}
As is seen, the renormalon model for the set of twist-4 DAs depends only on
one free parameter $\delta ^{2}$. It is related to the matrix element of the
local operator 
\[
\left\langle 0\left| \overline{d}\gamma _{\nu }ig\widetilde{G}_{\mu \rho
}u\right| \pi ^{+}(p)\right\rangle =\frac{1}{3}f_{\pi }\delta ^{2}\left[
p_{\rho }g_{\mu \nu }-p_{\mu }g_{\rho \nu }\right],
\]
\begin{equation}
\delta ^{2}(\mu _{0}^{2})\simeq 0.2\,\,\rm{GeV}^{2}  \label{eq:11}
\end{equation}
and estimated from various 2-point QCD sum rules \cite{NS94}.

In the case of the asymptotic DA, the pion twist-4 DAs were computed in 
Ref.\ \cite{BGG04}. In this paper we apply the results of Ref.\ \cite{BGG04} 
to a more general situation. To this end, we rewrite the leading twist DA 
(\ref{eq:8}) in the form 
\begin{equation}
\varphi _{\pi }(u,\,\mu _{F}^{2})=\varphi _{asy}(u)\sum_{n=0}^{\infty
}K_{n}(\mu _{F}^{2})u^{n}.  \label{eq:12}
\end{equation}%
This form is more suitable for calculations and leads to compact expressions
for the higher twist DAs. The DA $\varphi _{\pi }(u,\,\mu _{F}^{2})$ can
also be expanded over $\overline{u}$ with the same coefficients $K_{n}(\mu
_{F}^{2})$ and, hence, Eq.\ (\ref{eq:12}) preserves the symmetry of the
distribution amplitude under the replacement $u\leftrightarrow \overline{u}$%
, even if this is not explicitly seen from Eq.\ (\ref{eq:12}).

For DAs containing two nonasymptotic terms $C_{2}^{3/2}(u-\overline{u})$ and 
$C_{4}^{3/2}(u-\overline{u})$, the sum (\ref{eq:12}) \ runs over $n=0,1,..4$
and the coefficients $K_{n}(\mu _{F}^{2})$ are given by the equalities 
\[
K_{0}(\mu _{F}^{2})=1+6b_{2}(\mu _{F}^{2})+15b_{4}(\mu
_{F}^{2}),\,\,K_{1}(\mu _{F}^{2})=-30\left[ b_{2}(\mu _{F}^{2})+7b_{4}(\mu
_{F}^{2})\right], 
\]

\[
K_2(\mu _F^2)=30\left[ b_2(\mu _F^2)+28b_4(\mu _F^2)\right] ,\,\,K_3(\mu
_F^2)=-1260b_4(\mu _F^2),\, 
\]

\begin{equation}
\,K_{4}(\mu _{F}^{2})=630b_{4}(\mu _{F}^{2}).  \label{eq:13}
\end{equation}
Calculation of the three-particle DAs (\ref{eq:10}) is straightforward. The
two-particle DAs $\varphi _{1}^{(4)}(u,\mu _{F}^{2})$ and $\varphi
_{2}^{(4)}(u,\mu _{F}^{2})$ have the form
\[
\varphi _{1}^{(4)}(u,\mu _{F}^{2})=\sum_{n=0}^{4}K_{n}(\mu _{F}^{2})\varphi
_{n}^{1}(u),
\]
\begin{equation}
\varphi _{2}^{(4)}(u,\mu _{F}^{2})=\sum_{n=0}^{4}K_{n}(\mu _{F}^{2})\varphi
_{n}^{2}(u).  \label{eq:14}
\end{equation}%
Their components $\varphi _{n}^{1}(u)$ and $\varphi _{n}^{2}(u)$ are given
by the following expressions:
\[
\varphi _{0}^{1}(u)=\delta ^{2}\left\{ \overline{u}\left[ \ln \overline{u}-%
\rm{Li}_{2}(\overline{u})\right] +u\left[ \ln u-\rm{Li}_{2}(u)\right]
-u\overline{u}+\frac{\pi ^{2}}{6}\right\},
\]
\[
\varphi _{1}^{1}(u)=\delta ^{2}\left\{ \overline{u}\left[ \left( 1+\frac{
\overline{u}}{2}-\frac{\overline{u}^{2}}{3}\right) \ln \overline{u}-\rm{
Li}_{2}(\overline{u})\right] +u\left[ \left( 1+\frac{u}{2}-\frac{u^{2}}{3}
\right) \ln u-\rm{Li}_{2}(u)\right] \right. 
\]
\[
\left. -\frac{5}{6}u\overline{u}+\frac{1}{2}u^{2}\overline{u}^{2}+\frac{\pi
^{2}}{6}\right\}, 
\]
\[
\varphi _{2}^{1}(u)=\delta ^{2}\left\{ \overline{u}\left[ \left( 1+\overline{
u}-\frac{2}{3}\overline{u}^{2}\right) \ln \overline{u}-\rm{Li}_{2}(%
\overline{u})\right] +u\left[ \left( 1+u-\frac{2}{3}u^{2}\right) \ln u-
\rm{Li}_{2}(u)\right] \right. 
\]
\[
\left. -\frac{2}{3}u\overline{u}+\frac{5}{4}u^{2}\overline{u}^{2}+\frac{\pi
^{2}}{6}\right\}, 
\]
\[
\varphi _{3}^{1}(u)=\delta ^{2}\left\{ \overline{u}\left[ \left( 1+\frac{3}{2%
}\overline{u}-\frac{7}{6}\overline{u}^{2}+\frac{1}{4}\overline{u}^{3}-\frac{1%
}{10}\overline{u}^{4}\right) \ln \overline{u}-\rm{Li}_{2}(\overline{u})%
\right] \right. 
\]
\[
+u\left[ \left( 1+\frac{3}{2}u-\frac{7}{6}u^{2}+\frac{1}{4}u^{3}-\frac{1}{10}
u^{4}\right) \ln u-\rm{Li}_{2}(u)\right] 
\]
\[
\left. -\frac{31}{60}u\overline{u}+\frac{257}{120}u^{2}\overline{u}^{2}-
\frac{1}{3}u^{3}\overline{u}^{3}+\frac{\pi ^{2}}{6}\right\} ,
\]
\[
\varphi _{4}^{1}(u)=\delta ^{2}\left\{ \overline{u}\left[ \left( 1+2
\overline{u}-\frac{11}{6}\overline{u}^{2}+\frac{3}{4}\overline{u}^{3}-\frac{3
}{10}\overline{u}^{4}\right) \ln \overline{u}-\rm{Li}_{2}(\overline{u})
\right] \right. 
\]
\[
+u\left[ \left( 1+2u-\frac{11}{6}u^{2}+\frac{3}{4}u^{3}-\frac{3}{10}
u^{4}\right) \ln u-\rm{Li}_{2}(u)\right] 
\]
\begin{equation}
\left. -\frac{23}{60}u\overline{u}+\frac{47}{15}u^{2}\overline{u}^{2}-\frac{
61}{45}u^{3}\overline{u}^{3}+\frac{\pi ^{2}}{6}\right\},   \label{eq:15}
\end{equation}
and
\[
\varphi _{0}^{2}(u)=\delta ^{2}\left[ u^{2}\ln u+\overline{u}^{2}\ln 
\overline{u}+u\overline{u}\right],
\]
\[
\varphi _{1}^{2}(u)=\delta ^{2}\left[ u^{2}\ln u+\overline{u}^{2}\ln 
\overline{u}+u\overline{u}+\frac{1}{2}u^{2}\overline{u}^{2}\right],
\]
\[
\varphi _{2}^{2}(u)=\delta ^{2}\left[ u^{2}\ln u+\overline{u}^{2}\ln 
\overline{u}+u\overline{u}+\frac{5}{6}u^{2}\overline{u}^{2}\right],
\]
\[
\varphi _{3}^{2}(u)=\delta ^{2}\left[ u^{2}\ln u+\overline{u}^{2}\ln 
\overline{u}+u\overline{u}+\frac{13}{12}u^{2}\overline{u}^{2}-\frac{1}{6}
u^{3}\overline{u}^{3}\right],
\]
\begin{equation}
\varphi _{4}^{2}(u)=\delta ^{2}\left[ u^{2}\ln u+\overline{u}^{2}\ln 
\overline{u}+u\overline{u}+\frac{77}{60}u^{2}\overline{u}^{2}-\frac{13}{30}
u^{3}\overline{u}^{3}\right],  \label{eq:16}
\end{equation}
where $\rm{Li}_{a}(x)=\sum_{n=1}^{\infty }x^{n}/n^{a}$.

\begin{figure}[t]
\centering\epsfig{file=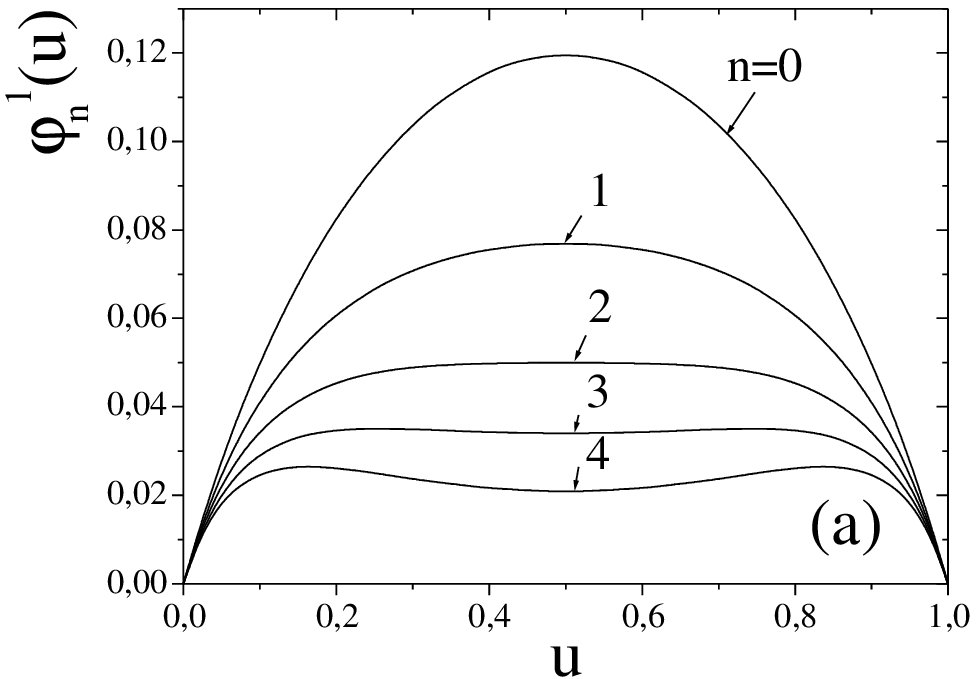,height=6.5cm,width=8.0cm,clip=}
\hspace{1.5cm}
\centering\epsfig{file=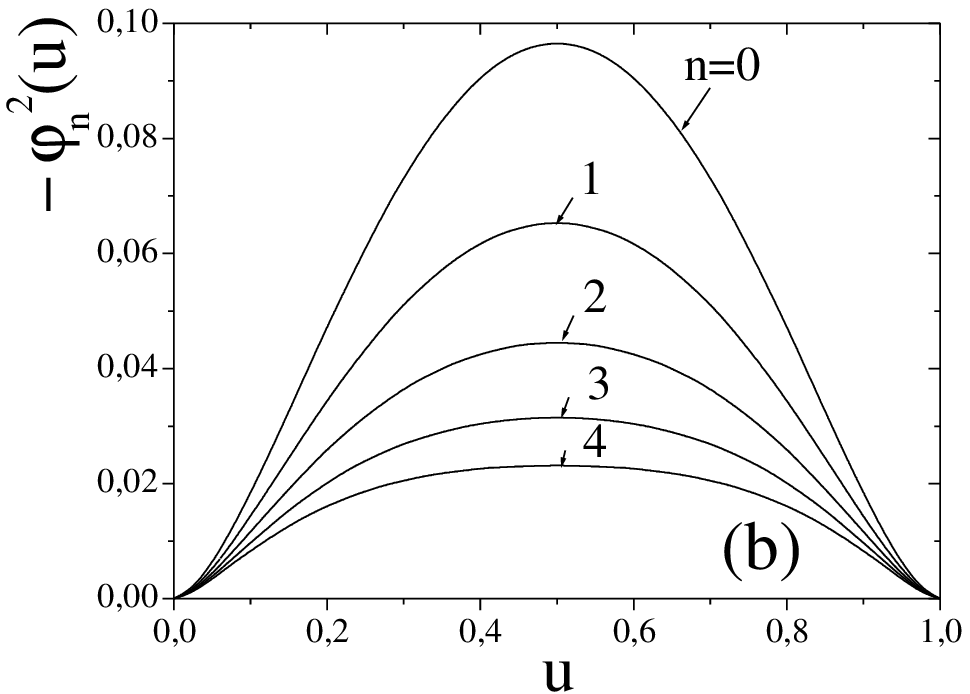,height=6.5cm,width=8.0cm,clip=}
\vspace{0.0cm}
\caption{ The components of the two-particle twist-4 DAs $\varphi _n^1(u)$ 
{\bf (a)} and $-\varphi _n^2(u)$ {\bf (b)} as functions of $u$. The
normalization constant is chosen equal to $\delta ^2=1$.}
\label{fig:brfig1}
\end{figure}

The functions $\varphi _{n}^{1}(u)$ and $\varphi _{n}^{2}(u)$ are shown in
Fig.\ \ref{fig:brfig1}. As is seen, the shapes of the functions $\varphi _{n}^{2}(u)$ are
identical to each other, difference being only in their normalization. On
the contrary, the functions $\varphi _{3}^{1}(u)$ and $\varphi _{4}^{1}(u)$
differ from $\varphi _{0(1,2)}^{1}(u)$ also in their shapes and have minima
at the point $u_{0}=1/2$ . With the constant $\delta ^{2}(\mu _{0}^{2})$
being fixed from the QCD sum rule, the twist-4 DAs $\varphi _{1}^{(4)}(u,\mu
_{F}^{2})$ and $\varphi _{2}^{(4)}(u,\mu _{F}^{2})$, as well as ones that
are given by Eq. (\ref{eq:10}) depend only on the parameters $b_{2}(\mu
_{0}^{2}),\,\,b_{4}(\mu _{0}^{2})$. In other words, in the framework of the
renormalon approach the twist-4 DAs of the pion are determined by the
function $\varphi _{\pi }(u,\mu _{F}^{2})$ unambiguously.

\section{The pion electromagnetic form factor within the QCD LCSR method}

In this section we apply the twist-4 DAs for calculation of the pion
electromagnetic form factor. We use the QCD LCSR method, which is one of the
powerful tools to estimate nonperturbative components of exclusive quanities 
\cite{BBK89}. The LCSR expression for the pion electromagnetic FF was
derived in Refs.\ \cite{BH94,Br00}. It was reanalyzed recently in Ref.\ \cite
{Bijnen}, where a sign error in the previous calculation of the twist-4
contribution to FF was corrected. Our approach to the twist-4 term leads to
further improvement of the prediction for FF, because the twist-4 DAs
obtained in the previous section encompass contributions arising from higher
conformal spins.

It is worth noting that the renormalon technique was successfully employed
for studying the light mesons electromagnetic and transition FFs \cite
{A01,A04,GK98}. In the works \cite{A01,A04}, the power-suppressed corrections
to these FFs were found using the running coupling method. The latter leads
to Borel resummed hard-scattering amplitudes of the relevant subprocesses
and necessitate calculation of the QCD factorization formulas applying the
principal value prescription. In Refs.\ \cite{A01,A04} it was demonstrated
that the running coupling method allows one to take into account both the
hard and soft components of the FFs.

The LCSR method is based on the analysis of the correlation function 
\begin{equation}  \label{eq:17}
T_{\mu \nu }(p,q)=i\int d^4xe^{iqx}\left\langle 0\left| T\left\{ j_\mu
^5(0)j_\nu ^{em}(x)\right\} \right| \pi ^{+}(p)\right\rangle,
\end{equation}
where $j_\mu ^5=\overline{d}\gamma _\mu \gamma _5u$ and $j_\nu ^{em}=e_u 
\overline{u}\gamma _\nu u+e_d\overline{d}\gamma _\nu d$ is the quark
electromagnetic current. The contribution of the pion intermediate state is
given by 
\begin{equation}  \label{eq:18}
T_{\mu \nu }(p,q)=2if_\pi (p-q)_\mu p_\nu F_\pi (Q^2)\frac 1{m_\pi
^2-(p-q)^2}.
\end{equation}
Here $f_\pi $ is the pion decay constant and $F_\pi (Q^2)$ is the pion
electromagnetic FF. Because $p^2=m_\pi ^2$ and $q^2=-Q^2$, the correlation
function (\ref{eq:17}) actually depends on one invariant variable $s=(p-q)^2$%
. For large negative values of $s$ and $q^2$, this correlator can be computed
in QCD. In the QCD sum rule method by matchig between the dispersion
relation in terms of contributions of hadronic states and the QCD
calculation at Euclidean momenta, one can estimate the hadronic quantities
under consideration, in our case, the pion FF $F_\pi (Q^2)$. This is the
common idea sharing by QCD sum rule methods, the difference being in
approaches to compute the correlation function (\ref{eq:17}) within QCD.

When the $q^2$ and $(p-q)^2$ are spacelike and large, the correlation
function can be expanded near the light-cone in terms of the pion DA of
increasing twist. As a result contributions to $F_\pi (Q^2)$ coming from the
pion DAs of different twists can be found. The leading twist (twist-2)
light-cone sum rule for $F_\pi (Q^2)$ is (hereafter $m_\pi ^2=0$) \cite{BH94}

\begin{equation}  \label{eq:19}
F_\pi ^{(2)}(Q^2)=\int_{u_0}^1du\varphi _\pi (u,\mu _F^2)\exp \left[ -\frac{
\overline{u}Q^2}{uM^2}\right] ,
\end{equation}
where 
\begin{equation}  \label{eq:20}
u_0=\frac{Q^2}{s_0+Q^2}.
\end{equation}
In Eqs.\ (\ref{eq:19}) and (\ref{eq:20}) $s_0$ is the duality interval; $M^2$
is the Borel variable.

The accuracy of the LCSR (\ref{eq:19}) was improved by calculating $O(\alpha
_{\rm{S}})$ correction to the twist-2 part, as well as including into
consideration twist-4 and twist-6 contributions \cite{Br00,Bijnen}. Finally,
the $F_\pi (Q^2)$ takes the following form: 
\begin{equation}  \label{eq:21}
F_\pi (Q^2)=F_\pi ^{(2)}(Q^2)+F_\pi ^{(2,\alpha _{\rm{S}})}(Q^2)+F_\pi
^{(4)}(Q^2)+F_\pi ^{(6)}(Q^2).
\end{equation}
The details of calculations and the explicit expression for $F_\pi
^{(2,\alpha \rm{_S})}(Q^2)$ can be found in Ref.\ \cite{Br00}. Here we
only remark that, namely, this contribution provides the standard QCD
asymptotics $\sim \alpha _{\rm{S}}/Q^2$ of the form factor.

The twist-4 term $F_\pi ^{(4)}(Q^2)$ is given by expression 
\begin{equation}  \label{eq:22}
F_\pi ^{(4)}(Q^2)=\int_{u_0}^1du\frac{\widetilde{\varphi }_4(u,\mu _F^2)}{%
uM^2}\exp \left[ -\frac{\overline{u}Q^2}{uM^2}\right] +\frac{u_0\widetilde{%
\varphi }_4(u_0,\mu _F^2)}{Q^2}e^{-s_0/M^2},
\end{equation}
where 
\begin{equation}  \label{eq:23}
\widetilde{\varphi }_4(u,\mu _F^2)=2u\left[ \frac d{du}\varphi
_2^{(4)}(u,\mu _F^2)-\overline{u}\frac{d^2}{du^2}\varphi _2^{(4)}(u,\mu _F^2)
\right].
\end{equation}
The difference between Eq.\ (\ref{eq:23}) and the relevant formula in Ref.\ 
\cite{Bijnen} is connected with the definition of the distribution amplitude 
$\varphi _2^{(4)}(u,\mu _F^2)$. In fact, the twist-4 DA $g_{2\pi }(u,\mu
_F^2)$ used in Ref. \cite{Bijnen} can be written in terms of $\varphi
_2^{(4)}(u,\mu _F^2)$
\[
g_{2\pi }(u,\mu _F^2)=\frac d{du}\varphi _2^{(4)}(u,\mu _F^2). 
\]

The factorizable twist-6 contribution to the LCSR was computed in Ref.\ \cite
{Br00} 
\begin{equation}  \label{eq:24}
F_\pi ^{(6)}(Q^2)=\frac{4\pi \alpha _{\rm{S}}(\mu _R^2)C_F}{3f_\pi ^2Q^4}
\left\langle 0\left| \overline{q}q\right| 0\right\rangle ^2,
\end{equation}
by means of the quark condensate density.

\section{Constraints on the pion DAs}

The LCSR expression for the pion electromagnetic FF and the twist-4 DAs
obtained in the framework of the renormalon approach can be used to extract
constraints on the input parameters $b_{2}(\mu _{0}^{2})$ and $b_{4}(\mu
_{0}^{2})$. In order to perform numerical computations, we need to fix values
of various parameters appearing in the relevant expressions. Namely , we
take the Borel parameter $M^{2}$ within the interval $0.8<M^{2}<1.5\,\,
\rm{GeV}^{2}$ and accept for the factorization and renormalization
scales the following value: 
\begin{equation}
\mu _{F}^{2}=\mu _{R}^{2}=\overline{u}Q^{2}+uM^{2}.  \label{eq:25}
\end{equation}
For the QCD coupling $\alpha \rm{_{S}}(\mu _{R}^{2})$ the two-loop
expression with $\Lambda _{3}=0.34\,\,\rm{GeV}$ is used. The value of
the duality parameter $s_{0}=0.7\,\,\rm{GeV}^{2}$ is borrowed from 
Shifman-Vainshtein-Zakharov
sum rule \cite{SVZ} for the correlator of two $\overline{u}\gamma _{\mu
}\gamma _{5}d$ currents. The normalization scale is set equal to $\mu
_{0}^{2}=1\;\rm{GeV}^{2}$.

\begin{figure}[t]
\centering\epsfig{file=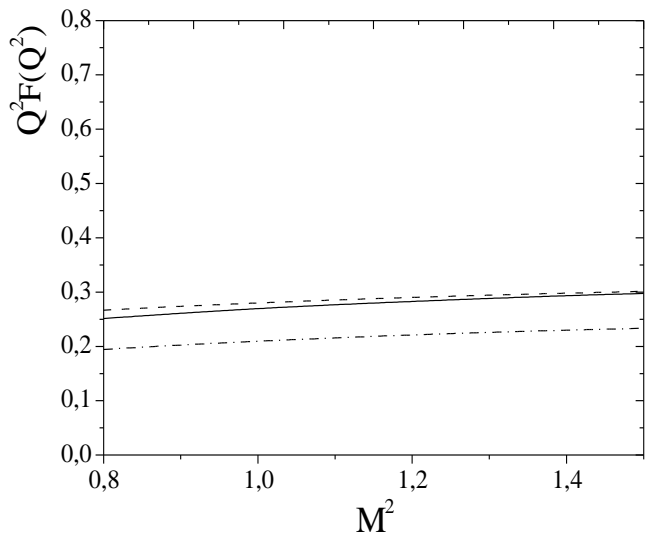,height=7cm,width=8.5cm,clip=}
\caption{
The dependence of the light-cone sum rule on the Borel parameter.
The asymptotic DA is used. For the solid curve $Q^{2}=1\,\rm{GeV}^{2}$
, for the dashed curve $Q^{2}=4\,\rm{GeV}^{2}$, and for the dot-dashed
one $Q^{2}=10\,\rm{GeV}^{2}$.}
\label{fig:brfig2}
\end{figure}

The Borel parameter dependence of the LCSR for different values of 
$Q^2$ is depicted in Fig.\ \ref{fig:brfig2}. From this figure one can conclude that
the prediction for the FF is rather stable in the exploring range of $M^2$.
In what follows we choose the Borel parameter equal to $M^2=1\,\,\rm{GeV}^2$.

\begin{figure}[t]
\centering\epsfig{file=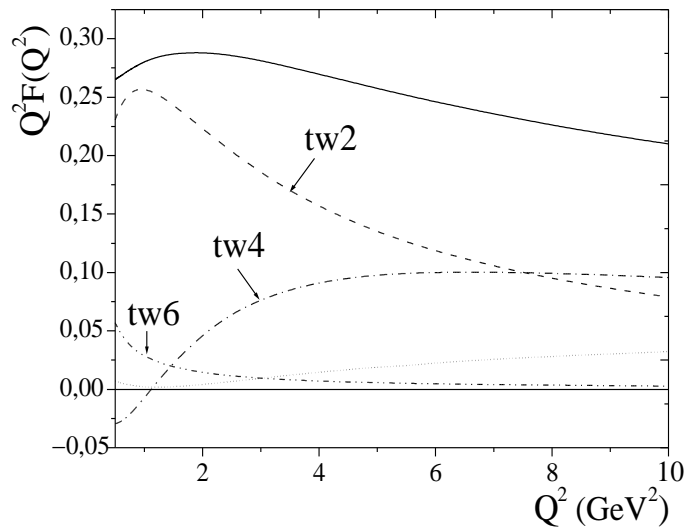,height=7cm,width=9cm,clip=}
\caption{
 The pion electromagnetic FF as a function of $Q^2$. The results are
obtained employing the asymptotic DA. The solid line corresponds to the sum
of the all contributions (\ref{eq:21}). The dotted line shows $O(\alpha _{
\rm{S}})$ correction to the twist-2 term.}
\label{fig:brfig3}
\end{figure}

The scaled FF $Q^{2}F_{\pi }(Q^{2})$ and its different components are
depicted in Fig.\ \ref{fig:brfig3}. In the calculations, the pion asymptotic DA and twist-4
DA $\varphi _{2}^{(4)}(u)$ obtained from the renormalon approach are used.
As is seen at $Q^{2}\simeq 7.5\,\rm{GeV}^{2}$, the twist-4 contribution
to the form factor exceeds the twist-2 one. This is important consequence of
the higher conformal spin (renormalon) effects containing in the DA $\varphi
_{2}^{(4)}(u)$. In Ref.\ \cite{Bijnen} the twist-4 term was calculated
employing the asymptotic, i.e. the lowest conformal spin, form of $g_{2\pi
}(u,\mu _{F}^{2})$. This form leads to the combination 
\begin{equation}
\widetilde{\varphi }_{4}(u,\mu _{F}^{2})=\frac{20}{3}\delta ^{2}(\mu
_{F}^{2})u\overline{u}\left[ 1-u(7-8u)\right],  \label{eq:26}
\end{equation}
with
\[
\delta ^{2}(\mu _{F}^{2})=\delta ^{2}(\mu _{0}^{2})\left[ \frac{\alpha _{
\rm{S}}(\mu _{F}^{2})}{\alpha _{\rm{S}}(\mu _{0}^{2})}\right]
^{8C_{F}/3\beta _{0}}. 
\]

\begin{figure}[tb]
\centering\epsfig{file=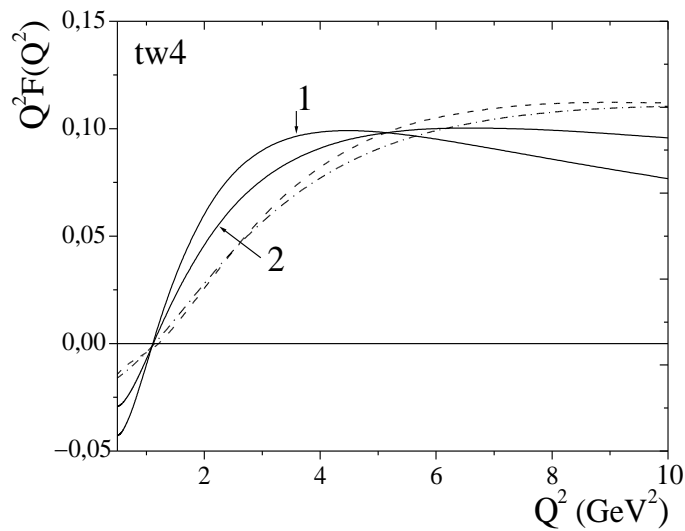,height=7cm,width=9cm,clip=}
\caption{
 The twist-4 term as a function of $Q^{2}$. The curve $1$ is computed
using the ordinary asymptotic twist-4 DA. The twist-4 DAs obtained employing
the renormalon method lead to the predictions shown by the line 2 and the
broken lines. The correspondence between the lines and the parameters 
$b_{2}(\mu _{0}^{2}),\,\,b_{4}(\mu _{0}^{2})$ is: the line $2$, $b_{2}(\mu
_{0}^{2})=0,\,\,b_{4}(\mu _{0}^{2})=0$; the dashed line, $b_{2}(\mu
_{0}^{2})=0.17,\,\,b_{4}(\mu _{0}^{2})=-0.05$; the dot-dashed line,  
$b_{2}(\mu _{0}^{2})=0.2,\,\,b_{4}(\mu _{0}^{2})=0$.}
\label{fig:brfig4}
\end{figure}

Comparing the twist-4 contributions found in the context of the different
methods, one reveals their interesting features. The corresponding predictions 
are plotted in Fig.\ \ref{fig:brfig4}. Here the curves $1$ and $2$ are
computed using the standard asymptotic DA (see Eq.\ (\ref{eq:26})) and the 
$\varphi _{2}^{(4)}(u)$ from the renormalon approach with $b_{2}(\mu
_{0}^{2})=b_{4}(\mu _{0}^{2})=0,$ respectively. The main difference between
them is that the higher conformal spin effects shift the maximum of the
twist-4 term towards larger values of $Q^{2}$. This feature of the twist-4 
term is more pronounced for DAs with $b_{2}(\mu _{0}^{2})\neq 0,$ $b_{4}(\mu
_{0}^{2})\neq 0$ (the broken lines in Fig. \ref{fig:brfig4}). Indeed, if the curve $2$ 
takes its maximal value at $Q_0^2 \simeq 6\ \rm{GeV}^2$,  for the broken 
lines we find $Q_0^2 \simeq 11\ \rm{GeV}^2$. Starting from $Q_0^2$, 
the twist-4 term slowly decreases, remaining larger than the standard 
prediction (the curve $1$). Such a modification of the asymptotic behavior is
another effect of the higher conformal spins.

Calculations of Ref.\ \cite{Br00,Bijnen} correspond essentially to the
''minimal'' model of the twist-4 effects, where the restriction to the
lowest conformal spin (a few lowest spins) probably underestimates the
effect, while the renormalon model is a ''maximal'' model, where these
effects are probably somewhat overestimated. Therefore, the renormalon model
for the twist-4 DAs allows us, for the first time, to put a theoretically
justified bound on the twist-4 contribution to the pion form factor. Actually 
the change in absolute value of the twist-4 correction is not too dramatic, 
as one might expect. Thus, the ratio

$$
\frac{tw4^{ren}(Q^2)}{tw4^{stand}(Q^2)}
$$
for the values $b_2(\mu_0^2)=b_4(\mu_0^2)=0$ is equal to $1.25$ at 
$Q^2=10\ \rm{GeV}^2$ and to $2.45$ at $Q^2=100\ \rm{GeV}^2$. 

\begin{figure}[tb]
\centering\epsfig{file=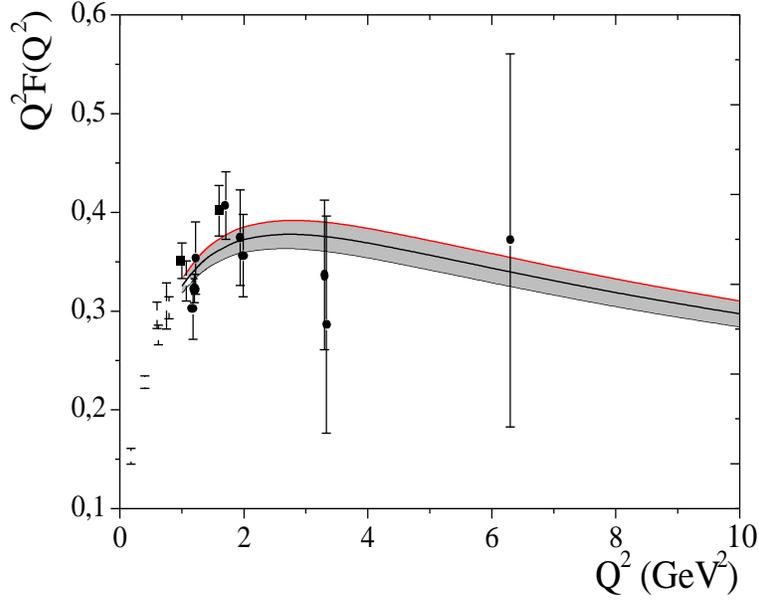,height=8cm,width=10cm,clip=}
\caption{
The $1\sigma $ region for the scaled pion FF $Q^{2}F_{\pi }(Q^{2})$.
The data are taken from Refs.\ \cite{Beb} (the circles) and \cite{Vol} (the
rectangles). In the $1\sigma $ analysis only the solid data points are used.
For the central solid curve the input parameters are $b_{2}(\mu
_{0}^{2})=0.2,\,\,b_{4}(\mu _{0}^{2})=0$.}
\label{fig:brfig5}
\end{figure}

In the renormalon approach, the twist-4 DAs and, hence, the twist-4
contribution to the form factor, depends on the input parameters and is not a
constant background for the leading twist contribution. Therefore, performed 
$1\sigma $ analysis results in conclusions, which differ from those made in
Ref.\ \cite{Bijnen}. In Fig.\ \ref{fig:brfig5} we demonstrate the results of such $1\sigma$ analysis. 
The data points included into the fitting procedure are shown by
the solid points. Here we take into account the data $Q^{2}\geq 1.18\,\,
\rm{GeV}^{2}$ reported in Ref.\ \cite{Beb}, and two new data points at $%
Q^{2}=1;\,\,1.6\,\,\rm{GeV}^{2}$ obtained by the $F_{\pi }$
collaboration \cite{Vol}. From this analysis we extract the value of the
input parameter $b_{2}(1\,\rm{GeV}^{2})$, in the case of the DA with
one nonasymptotic term 
\begin{equation}
b_{2}(1\ \rm{GeV}^{2})=0.2\pm 0.03.  \label{eq:27}
\end{equation}
For the pion DA with two nonasymptotic terms, we get 
\begin{equation}
b_{2}(1\ \rm{GeV}^{2})=0.2\pm 0.03,\,\,b_{4}(1\ \rm{GeV}
^{2})=-0.03\pm 0.06.  \label{eq:28}
\end{equation}
From Eqs.\ (\ref{eq:27}) and (\ref{eq:28}) it becomes evident that impact of $b_{2}(\mu
_{0}^{2})$ on the numerical computations is more important than a role
played by $b_{4}(\mu _{0}^{2})$. Our analysis does not exclude also DAs with
two positive input parameters. But it is worth noting that in our
consideration we have used the data \cite{Beb} which were extracted
indirectly from the pion electroproduction experiments through a
model-dependent extrapolation to the pion pole. Moreover, the points 
$Q^{2}>2\ \rm{GeV}^{2}$ are imprecise suffering from the large errors
and they are rather sparse. Correct and direct measurements of the form
factor at $Q^{2}\geq 2\ \rm{GeV}^{2}$ will improve the performed
analysis and allow one to put more strong constraints on the pion DAs.

In the region $1\ \rm{GeV}^{2}\leq $ $Q^{2}<10\ \rm{GeV}^{2}$, our
prediction for the pion electromagnetic FF can be fitted to the following
formula:

\begin{equation}
Q^{2}F_{\pi }(Q^{2})=(0.2227\pm 0.011)+\frac{0.5107\pm 0.0115}{Q^{2}}-\frac{%
0.4284\pm 0.0165}{Q^{4}},  \label{eq:29}
\end{equation}%
where the uncertainties in the numerical coefficients are connected with the
experimental errors.

\begin{figure}[tb]
\centering\epsfig{file=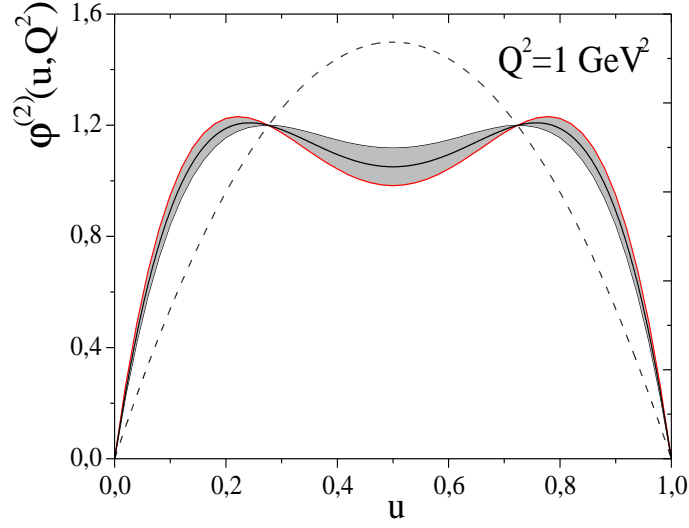,height=7cm,width=9cm,clip=}
\caption{
The pion leading twist DA extracted in this work. The scale $Q^{2}$
is fixed at $1\,\,\rm{GeV}^{2}$. For the central solid curve $b_{2}(\mu
_{0}^{2})=0.2$. For comparison the asymptotic distribution amplitude is also
shown (dashed curve).}
\label{fig:brfig6}
\end{figure}

\begin{figure}[t]
\centering\epsfig{file=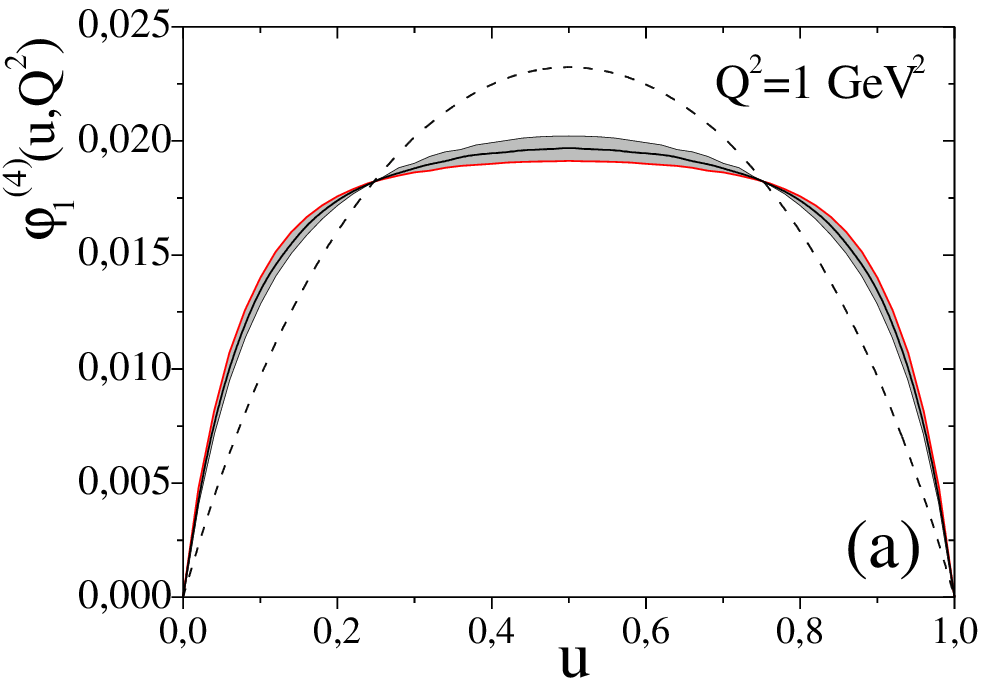,height=6.5cm,width=8.0cm,clip=}
\hspace{1.5cm}
\centering\epsfig{file=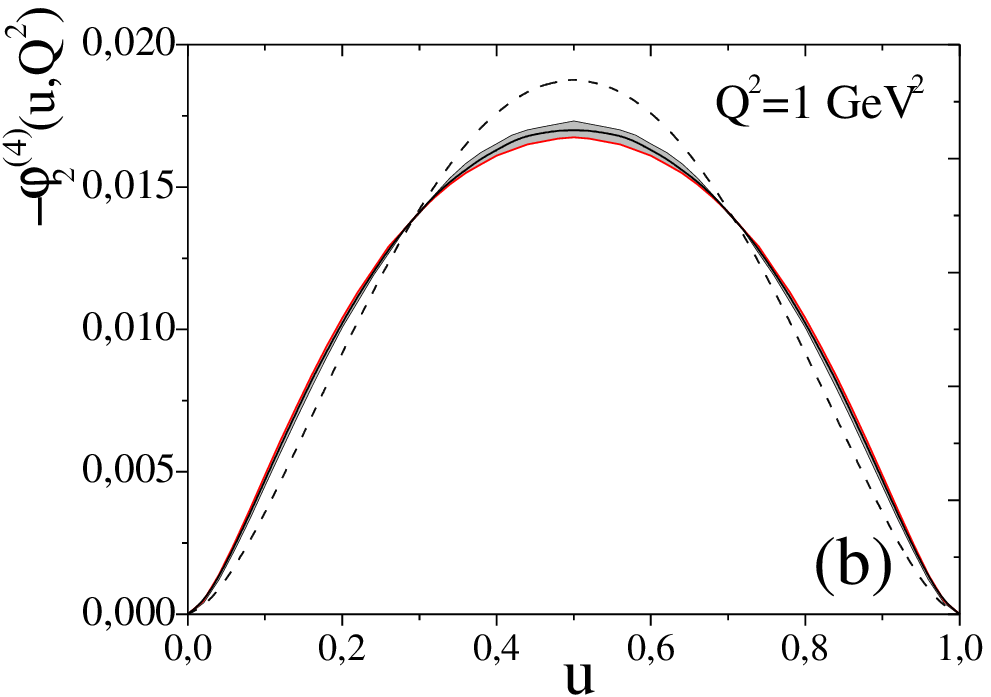,height=6.5cm,width=8.0cm,clip=}
\vspace{0.0cm}
\caption{ 
The two-particle twist-4 DAs $\varphi _{1}^{(4)}(u,Q^{2})$ 
{\bf (a)}, and $-\varphi _{2}^{(4)}(u,Q^{2})$ {\bf (b)} obtained within the
renormalon approach and using the constraint on the parameter $b_{2}(\mu
_{0}^{2})$ (\ref{eq:27}). By dashed lines, for comparison, we plot the DAs
obtained also within the renormalon approach, but using the pion asymptotic
leading twist DA.}
\label{fig:brfig7}
\end{figure}

The pion twist-2 and two-particle twist-4 DAs calculated using the parameter
(\ref{eq:27}) are shown in Figs.\ \ref{fig:brfig6} and \ref{fig:brfig7}. 
The shaded areas in the figures
are obtained varying the parameter $b_{2}(1\,\rm{GeV}^{2})$ within the
allowed interval. The pion leading twist DA in the middle point $u_0=0.5$
takes the values

\begin{equation}
\label{eq:30}
\varphi^{(2)}(0.5, 1\ \rm{GeV}^2)=1.05 \mp 0.07,\
\varphi^{(2)}(0.5, 10\ \rm{GeV}^2)=1.18 \mp 0.05.
\end{equation} 
This estimate is rather precise and does not contradict to the old 
Braun-Filyanov result,
$$
\varphi^{(2)}(0.5)=1.2 \pm 0.2,
$$ 
from the second paper in Ref.\ \cite{BBK89}.
The model DAs corresponding to Eq.\ (\ref{eq:28}) have the
similar behavior and are not shown in the figures.

\section{Concluding remarks}

In the present work we have used the renormalon approach to determine the
pion twist-4 DAs. In this approach the higher twist DAs are expressed in
terms of the leading twist DA unambiguously. This fact has allowed us to
avoid expansion of the higher twist DAs over the conformal spin and, at the
same time, to take into account higher conformal spin effects. Of course,
the renormalon approach is not suitable to model higher spin effects in the
leading twist DA. Nevertheless, it considerably restricts a possible form of
the higher twist DAs.

The obtained model DAs have been employed for computation of the pion
electromagnetic FF within the QCD LCSR method. For this purpose the correct
expression for the twist-4 contribution to $F_\pi (Q^2)$ has been used \cite
{Bijnen} and from comparison with the available experimental data the
constraints on the input parameters $b_2(\mu _0^2),$ $b_4(\mu _0^2)$ at $\mu
_0^2=1\ \rm{GeV}^2$ have been deduced.

The pion twist-2 DA $\varphi _{\pi }(u,Q^{2})$ was an object of numerous
investigations. It was modeled using the various theoretical schemes and
exclusive processes (see, for example, Refs.\ \cite{CZ84,A01,pion}). The
models found in the present work are close to ones predicted in Ref.\ \cite%
{A01}. In Ref.\ \cite{A01} the power-suppressed corrections to $F_{\pi
}(Q^{2})$ were evaluated in the framework of the standard hard-scattering
approximation and the running coupling method, which resulted in the Borel
resummed FF $[Q^{2}F_{\pi }(Q^{2})]^{res}$ , whereas in the present paper we
have computed the twist-4 contribution to $F_{\pi }(Q^{2})$ in the context
of the LCSR method and the renormalon-inspired twist-4 DAs. The new
contribution of this work is that the renormalon approach has allowed to put
an upper bound on the twist-4 contribution to the sum rules and obtain
estimates, for the first time, of the effects due to higher conformal spins.
We have gotten similar values of the pion DA parameters compared to other
studies, so the LCSR approach seems to be protected from large uncertainties
coming from higher twist corrections.

\bigskip\ 

{\bf Acknowledgments.}

\bigskip

I am grateful to Prof.\ V. M. Braun for drawing to my attention this problem,
reading the manuscript and for valuable comments. I would like also to thank
Prof.\ J. P. Blaizot and ''European Centre for Theoretical Studies in Nuclear
Physics and Related Areas'' (ECT*) for hospitality in Trento, where this
work was started and Prof.\ S. Randjbar-Daemi and members of the High Energy
Section in the A. Salam ICTP, where it was completed.

\end{document}